# Examining Passenger Vehicle Miles Traveled and Carbon Emissions in the Boston Metropolitan Area

**Tigran Aslanyan and Shan Jiang**

## Abstract

With spatial analytic, econometric, and visualization tools, this book chapter investigates greenhouse gas emissions for the on-road passenger vehicle transport sector in the Boston metropolitan area in 2014. It compares greenhouse gas emission estimations from both the production-based and consumption-based perspectives with two large-scale administrative datasets—the vehicle odometer readings (from individual vehicle annual inspection) and the road inventory data (containing road-segment level geospatial and traffic information). Based on spatial econometric models that examine socioeconomic and built environment factors contributing to the vehicle miles traveled (VMT) at the census tract level, it offers insights to help cities reduce VMT and carbon footprint for passenger vehicle travel. Finally, it recommends a pathway for cities and towns in the Boston metropolitan area to curb VMT and mitigate carbon emissions to achieve climate goals of carbon neutrality.

## Keywords

Vehicle miles traveled, carbon emissions, land use, transport, spatial econometrics

T. Aslanyan
Department of Urban and Environmental Policy and Planning, Tufts University
97 Talbot Avenue, Medford, MA 02155 USA
e-mail: tigran.aslanyan@tufts.edu

S. Jiang (corresponding author)
Department of Urban and Environmental Policy and Planning, Tufts University
97 Talbot Avenue, Medford, MA 02155 USA
e-mail: shan.jiang@tufts.edu
ORCID: 0000-0002-3483-5132

# 1 Introduction

Cities have become the center of social and economic activities, bringing together diverse populations and innovations for future growth and prosperity. As cities continue to grow exponentially—projected to contain two-thirds of the world's population by 2050—they become primary contributors of greenhouse gas (GHG) emissions across the globe and more vulnerable to environmental changes. Today, 70% of the world's energy-related GHG emissions come from cities (*C40* 2018); around one-third of the emissions in cities come from the transportation sector, 80% of which stem from on-road vehicle emissions. The negative externalities from personal automobiles, such as traffic congestion, carbon emissions, and traffic accidents, have been observed for a long time. Many policies have been implemented to mitigate the impacts of these externalities (Cook et al. 2014). Cities around the United States have released climate action plans, frameworks for measuring, planning, reducing GHG emissions, and responding to the climate crisis in the upcoming years. Government agencies and organizations have established short- and long-term goals based on climate analysis studies addressing crucial mobility and environmental problems. Preparation of an emissions inventory for the passenger transport sector serves as the starting point for developing comprehensive emission reduction strategies. However, methodologies used to date vary significantly, limiting government agencies' ability to compare and aggregate local, subnational, and national GHG emissions. Access to robust and high-quality data on GHG emissions is the key, as it could enable government agencies to evaluate the risks and opportunities of the issue (BSI 2013). A standard for measuring emissions is necessary at a state level if cities aim to promote collaborative efforts to address the environmental challenge.

This book chapter investigates the GHG emissions generated by on-road passenger vehicles in Massachusetts by taking advantage of two large administrative datasets – the annual vehicle registration and inspection records (from 2009 to 2014) and the road inventory data (MassDOT 2018). We compare the former estimates with the latter, which contain the annual average daily traffic (AADT) generated by the Highway Performance Monitoring System (FHWA, 2018), commonly used by transportation planners. The study findings will allow cities and towns to compare their emissions using these two approaches and perspectives (i.e., consumption-based and production-based) and assess alternative policies suitable for their future development.

Researchers, planners, and policymakers must act together to design effective policies to reduce vehicle miles traveled (VMT) and mitigate GHG emissions for the transportation sector. Toward this end, we develop a set of econometric models to examine the relationship between VMT, socioeconomic, and built environment determinants of the driving demand. We organize the chapter as follows. Section 2 reviews the literature on VMT estimation methods and determinants. Section 3 describes the methods and data used to estimate GHG emissions for passenger vehicles and spatial econometric models to explain factors that drive VMT in the study area. Section 4 discusses findings and results from the GHG emission estimations and the VMT models. In the final section, we summarize the limitations of this study and provide policy recommendations and suggestions for cities in the Boston metropolitan area to mitigate VMT and prepare sustainable transportation policies for a carbon-neutral future.

## 2 Literature Review

### *2.1 VMT Estimation Methods*

Vehicle miles traveled (VMT) is a widely used measure in transportation planning to allocate resources, estimate transportation emissions, assess traffic impacts, and adjust travel demand forecasts (Williams et al. 2016). It measures the typical amount of total vehicle travel in a geographic region over one year. However, it has been challenging to collect travel data for individual vehicles; therefore, various methodologies have been developed and applied to obtain VMT estimates (Kumapley and Fricker 1996, Fan et al. 2019).

There are two approaches to estimate VMT: traffic count-based and non-traffic count-based. The former method employs manual counts or sensors on roadway sections for a period. It then converts the data into annual counts using seasonal and time-of-day expansion factors (Williams et al. 2016). While the latter approach is most commonly used and simple to implement, it is typically designed for statewide use and may not be statistically valid for smaller geographic units of analysis (Liu et al. 2006). The most frequently adopted method by the Environmental Protection Agency (EPA 2015) is the Federal Highway Administration's (FHWA) traffic count-based Highway Performance Monitoring Systems (HPMS) to estimate VMT throughout the roads and highways in the U.S. (Kumapley and Fricker 1996).

On the other hand, the non-traffic-count-based VMT estimation approaches use socioeconomic data (including fuel sales, trip-making behavior, household size, household income, population, number of licensed drivers, and employment) to estimate VMT. However, the data collection can be costly, and historical data have often been employed to derive growth factors for projecting future VMT estimates. Three different types of data have been widely used, including household travel surveys (Handy et al. 2005, Huang et al. 2019), fuel-sales (Williams et al. 2016), and vehicle-inspection (odometer-reading) data (Reardon et al. 2016, Diao and Ferreira 2014). More recently, big data (such as GPS-based, or cell phone data) have become available to estimate VMT but may need traditional data sources and models for validation (Fan et al. 2019, NASEM 2019).

Some studies aim at investigating causation (Handy et al. 2005, Small and Van Dender 2007, Duranton and Turner 2011), while others look for the spatial and temporal association and correlation among variables (Cook et al. 2014, Diao and Ferreira 2014, Glass et al. 2012). Econometric models of VMT vary from ordinary least squares (OLS) regressions to three-stage least squares with panel data to spatial-lag and spatial-error models with pooled cross-sectional data.

### *2.2 Determinants of VMT*

Experts in transportation planning, engineering, and economics have tried to understand and quantify the relationships between key variables in travel behavior and VMT, and findings from the literature have been applied to various levels of policymaking. However, no consensus has been achieved on the relative magnitudes of the relationship between VMT and its determinants.

Empirical studies concentrated on the induced vehicle usage from an increase in highway length or fuel-efficiency standards find that changes in VMT are determined mainly by fuel cost, income, and specific urban characteristics (Cervero and Hansen 2002, Small and Van Dender 2007, Turner and Duranton 2009, Hymel 2014, Glass et al. 2012). However, the effect between income and VMT may not always be significant (Cook et al. 2014). Demographic variables (such as population growth, age, ethnicity, unemployment rate, and household size) are found to have significant effects on VMT. Built environment variables, such as land uses and public transit availability, also play important roles for VMT.

Both aggregate and disaggregate analyses have been conducted. The former includes using data at the neighborhood or fine-grained grid level (Diao and Ferreira 2014). The latter relies on micro individual-level characteristics which generally are obtained from household surveys (McMullen and Eckstein, 2013; Huang et al., 2019). McMullen and Eckstein (2013) conclude that higher lane miles, income, and employment in construction and public sectors lead to increased VMT per capita, while fuel price, transit use, and urban population density show an inverse association with travel demand. Diao and Ferreira (2014) find that one standard deviation variation in built-environment factors is associated with around 5,000 VMT differences in annual VMT per-household.

In this paper, we take advantage of the odometer readings data collected during vehicle inspections in Massachusetts to examine on-road passenger vehicle GHG emissions and determinants of VMT and compare this approach with traditional traffic-count based VMT and GHG emission estimations. We show the significant spatial variations in GHG emission estimations with these different approaches, which can be used as alternative evidence by cities when designing responsive policies to tackle climate change challenges.

## 3 Methodology and Data

This study examines the Boston metropolitan area governed by the Metropolitan Area Planning Council (MAPC), consisting of 22 cities and 79 towns and 699 census tracts. Massachusetts, where Boston is located, consists of 14 counties, 13 regional planning organizations, and 1478 census tracts (Fig. 1).

We estimate carbon dioxide emissions from passenger vehicles using two approaches: (1) with the VMT estimates from the vehicle inspection data, and (2) with the annual average daily traffic (AADT) data from the road inventory dataset and corresponding emission factors at the road segment level. We use the Massachusetts Vehicle Census dataset (MAPC, 2014) that includes all registered vehicles in Massachusetts with VMT estimates to accurately reflect the driving demand and trends of specific geographic localities.

**Fig. 1 Study area and regional planning organization boundaries in MA**

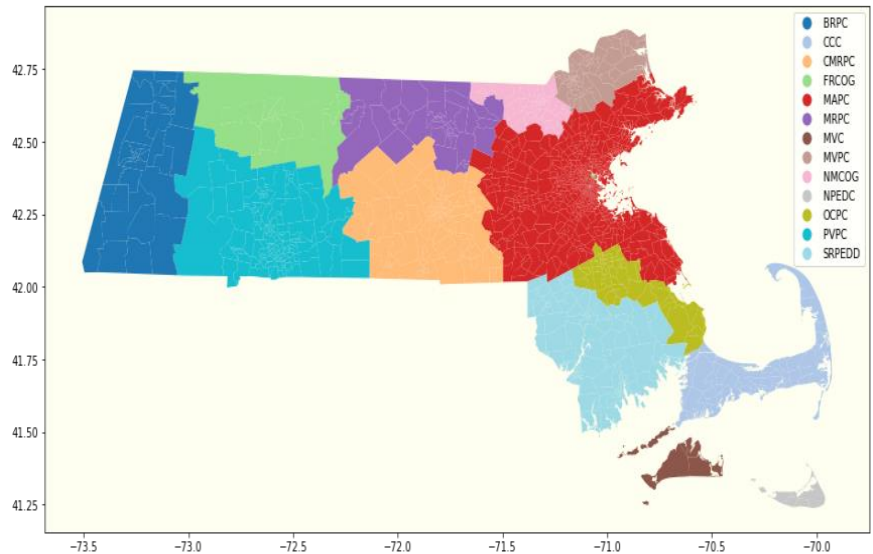

From the literature and our analyses, we find that reducing VMT is the key to mitigating GHG emissions. We then develop a set of econometric models to understand factors that influence VMT in the study area. These factors include the fuel economy, transportation accessibility, and key built environment variables (discussed in detail in Section 3.2.2). By incorporating journey-to-work travel mode variables, including carpooling and biking, we identify the extent to which these modes can substitute passenger vehicle travels. We discuss four model specifications with ordinary least squares (OLS), spatial lag model (SLM), spatial error model (SEM), and the mixed spatial error and lag model (SELM). The spatial analysis and spatial econometrics modeling introduced in this paper are all conducted using open-sourced Python libraries GeoPandas and PySAL (Rey et al., 2010).

## *3.1 Estimating GHG Emissions*

### 3.1.1 Data on VMT and Emission Factors

**Vehicle Inspection Data** Massachusetts Vehicles Census data contains every vehicle registered in MA between 2009 and 2014, with information from vehicle registrations, inspection records, and mileage ratings, published quarterly by the Metropolitan Area Planning Council (MAPC). The annual VMT estimates are based on the first and last odometer readings of each vehicle, making the data more reliable and accurate than other estimates from household travel surveys, which usually had small sample sizes. We select the anonymized summary data by census tracts and block groups for each calendar quarter with attributes including estimated vehicle age, fuel efficiency, and average daily vehicle miles traveled (DVMT) (Fig. 2).

**Fig. 2 Average Daily VMT per Passenger Vehicle in Massachusetts**

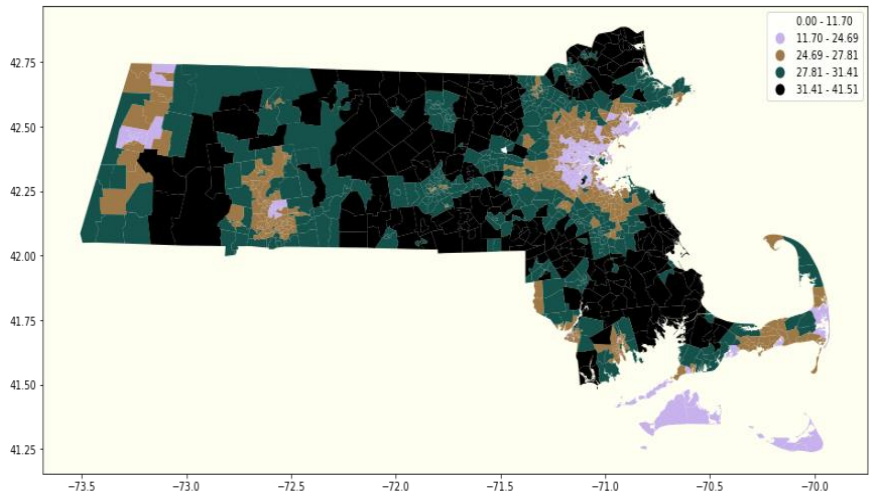

Data source: Massachusetts vehicles census data, 2014

**Road Inventory Data** For comparison purposes, in addition to the VMT estimate from the vehicle inspection data (i.e., 2014 Vehicle Census), we also obtain the road inventory data (MassDOT 2018). The latter contains geospatial information of every road segment in Massachusetts, their associated speed limit, functional class, road segment length, and annual average daily traffic (AADT) (see Fig 3). Based on the road 'Functional Class' attribute and data provided by the Boston metropolitan planning organization (MPO)—Central Transportation Planning Staff (CTPS), we apply a factor of 0.9266 for passenger VMT (and 0.0734 for freight VMT) for road segment with types of Interstate, principal arterial, minor arterial, and major collector. For the rest of the road types, we assume the VMT is 100% passenger VMT.

**Fig. 3 Annual Average Daily Traffic by Road Segment in Massachusetts**

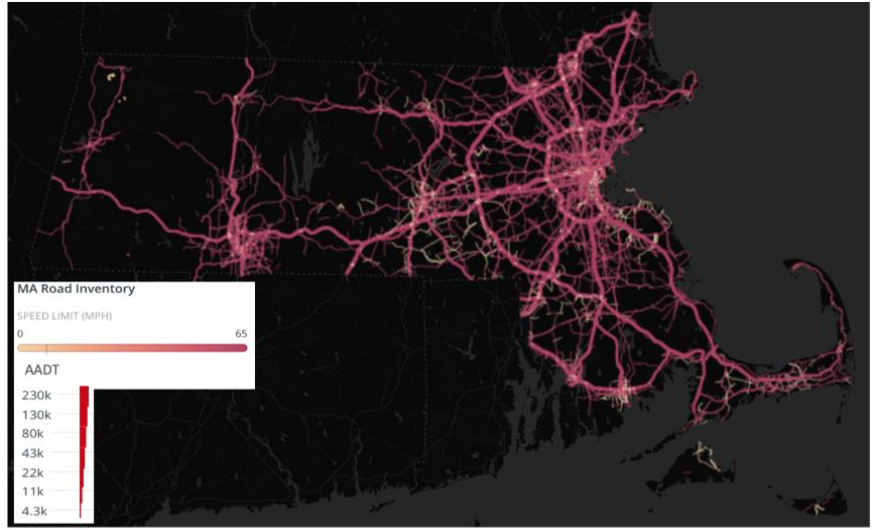

Data source: MassDOT Road Inventory data, 2018

**Emission Factor Data** For emission calculations, we use the EMFAC database by California Air Resources Board for 2014, which consists of the emission factors (EFs) for each speed bin, fuel type, and passenger vehicle type. To calculate the emission factor $EF(s,t)$ for each speed limit bin($s$) and each time of day($t$) by census tract ($ct$) or road segment (r), which are employed by the following two approaches discussed in Section 3.1.2 and 3.1.3, we adopt both the summary statistics of VMT distribution by speed and time of day (estimated by CTPS), and the weighted emission factor by speed ($EF_w$) estimated by AECOM (2016) based on the EMFAC database (see Table 1).

**Table 1 (a)VMT proportion by speed limit for different time of day, and (b) weighted emission factor (EF) by speed limit in the Boston metropolitan Area**

| Speed Bin (mph) | (a) VMT Proportion for time of day ($p_s(t)$) | | | | (b) Passenger Weighted EF ($EF_w$:gCO2e/mile) |
|---|---|---|---|---|---|
| | AM | MD | PM | NT | |
| (0, 5) | 2.30% | 9.13% | 3.33% | 3.93% | 1184.21 |
| [5, 10) | 0.43% | 0.57% | 0.38% | 0.20% | 1184.21 |
| [10, 15) | 3.16% | 4.47% | 4.03% | 1.38% | 873.37 |
| [15, 20) | 12.03% | 15.30% | 13.54% | 9.23% | 675.03 |
| [20, 25) | 17.01% | 18.92% | 17.31% | 13.04% | 539.09 |
| [25, 30) | 12.80% | 12.48% | 13.13% | 13.20% | 446.67 |
| [30, 35) | 10.95% | 9.16% | 10.46% | 13.10% | 383.63 |
| [35, 40) | 8.81% | 7.29% | 9.94% | 9.23% | 344.23 |
| [40,45) | 5.72% | 6.65% | 8.43% | 4.42% | 319.98 |
| [45,50) | 7.04% | 4.86% | 6.79% | 7.89% | 308.42 |
| [50,55) | 6.29% | 2.43% | 4.53% | 7.21% | 308.34 |
| [55,60) | 4.45% | 4.00% | 3.75% | 5.55% | 319.60 |
| [60,65) | 6.79% | 3.47% | 2.89% | 7.99% | 340.58 |
| [65, 65+) | 2.23% | 1.26% | 1.49% | 3.62% | 380.39 |
| | 100% | 100% | 100% | 100% | |

Note: (a) VMT proportion distribution by speed limits are estimated by CTPS, 2013. Total VMT by time of day are 16.62% for AM, 21.09% for PM, 32.66% for mid-day (MD), and 29.43% for nighttime (NT). (b)Weighted EFs by speed limits are synthesized by AECOM based on the EMFAC database.

As the distribution of road speed limits is different (across census tracts or road segments) during a different time of the day, we calculate the weighted emission factors by the speed limit and by the time of day, $EF(s,t)$, as follows: $EF(s,t) = \sum^s EF_w * p_s(t) / \sum^s p_s(t)$, where $EF_w$ is weighted EF by speed, and $p_s(t)$ is the VMT proportion by each speed and time of day discussed in Table 1.

The road inventory dataset includes the speed limit for each road segment. Depending on the two different approaches presented in the following sections, we extract the average road speed limit by census tract (in Section 3.1.2) and by road segment (in Section 3.1.3) and estimate corresponding EFs. We assume that all cars drive at or below the road speed limit. We calculate the corresponding weighted EFs by averaging EFs below the speed limit for each time period weighted by VMT proportions reported in Table 1. For example, to calculate the weighted EF in AM hours for a road segment with a speed limit of 30, we sum the products of VMT proportion by speed bins of (0,5], [5,10), [10,15), [15,20), [20,25), [25, 30) in the AM period and its corresponding passenger weighted EFs, and divide that value by the sum of the VMT proportions by these speed bins. The final weighted EFs by the speed limit and by time of day are presented in Table 2. These EFs will be employed for the GHG emission estimations in the following sections.

**Table 2 Emission factors (EF) by speed limit and time of day in the Boston metropolitan area, 2013**

| Time of Day | Average Speed (mph) | Weighted EF (gCO2e/mile) | Time of Day | Average Speed (mph) | Weighted EF (gCO2e/mile) |
|---|---|---|---|---|---|
| AM | 30 | 565.8 | MD | 30 | 642.69 |
| | 35 | 536.87 | | 35 | 614.55 |
| | 40 | 519.93 | | 40 | 591.23 |
| | 50 | 487.36 | | 50 | 568.65 |
| | 60 | 469.52 | | 60 | 550.55 |
| PM | 30 | 583.22 | NT | 30 | 559.88 |
| | 35 | 550.29 | | 35 | 528.45 |

| | | | |
|---|---|---|---|
| 40 | 526.19 | 40 | 514.83 |
| 50 | 499.35 | 50 | 477.2 |
| 60 | 487.85 | 60 | 456.8 |

### 3.1.2 Estimating GHG Emissions with Vehicle Inspection Data

We employ the following equations (1 and 2) to calculate the GHG emissions at the census tract level using the vehicle inspection data from the latest publicly available Vehicle Census dataset for Massachusetts in 2014. This dataset includes quarterly average daily passenger VMT, number of vehicles per census tract. We also estimate emission factors as discussed in Section 3.1.1.

$$E_{ct}^{VC} = \sum_{t} EF(s_{ct}, t) * VMT_{ct}^{y}(t) \qquad (1)$$

$$VMT_{ct}^{y}(t) = \sum_{q=1}^{4} \left\{ DVMT_{ct}(q, t) * PV_{ct}(q) * \left(\frac{365}{4} \, days\right) \right\} \qquad (2)$$

Where $E_{ct}^{VC}$ is total annual CO$_2$e emissions from passenger vehicles in each census tract (estimated with the odometer readings of the vehicle census data). $VMT_{ct}^{y}(t)$ is the total annual VMT in each census tract (*ct*). $DVMT_{ct}(q, t)$ is the average daily passenger VMT in each quarter *(q)* by the time of day *(t)* for census tract *(ct)*. VMT by the time-of-day distributions are provided by CTPS: 16.62% for AM, 21.09% for PM, 32.66% for mid-day (MD), and 29.43% for nighttime (NT). $PV_{ct}(q)$ is the number of vehicles in each census tract (*ct*) during each quarter (*q*). The reason for treating VMT by the different time of the day is that emission factors vary by time. $EF(s_{ct}, t)$ is the weighted emission factor by an average speed limit in each census tract ($s_{ct}$) for the time-of-day *t* (AM, PM, MD, NT) as discussed in Section 3.1.1. The average speed limit for each census tract is derived by the spatial analysis overlaying the road network layer with the census tract layer.

Based on Table 2 and the equations (1 and 2) listed above, we could see that the estimates generated here are in the higher bound, given that emission factors are calculated based on the vehicle registration's home census tracts. As residential areas usually have relatively lower speed limits than those in areas with major roads where on-road vehicles travel more frequently, the EFs in these areas are generally higher.

### 3.1.3 Estimating GHG with Road Inventory Data

To compare the GHG emissions estimated for vehicle owners' residential location (using the vehicle inspection data in Section 3.1.2.) with those generated on the road systems, here we use the road inventory data to derive the GHG emissions. We incorporate the road inventory data which includes AADT (collected by highway performance monitoring systems), road segment lengths, average speed limits, and the functional classifications.

$$E_{ct}^{RI} = \sum_r \sum_t EF(s_r,t) * AVMT_{r\,ct}^{y}(t) \quad (3)$$
$$AVMT_{r\,ct}^{y}(t) = AADT_{r\,ct}(t) * L_{r\,ct} * 365\ days \quad (4)$$

Where $E_{ct}^{RI}$ is the total annual CO$_2$e emissions in each census tract (*ct*) using the road inventory dataset. $AVMT_{r\,ct}^{y}(t)$ is the annual VMT for each road segment *r* within each census tract (*ct*) and for each time-of-day period *t* (AM, PM, MD, NT). $L_r$ is the length of each road segment. $EF(s_r)$ is the weighted average emission factor by road speed limit by the time of day (*t*) for road the segment $s_r$. The EF estimates are discussed in Section 3.1.1.

## *3.2 Examining Factors Contributing to VMT*

### **3.2.1 OLS, Spatial Lag, and Spatial Error Models**

First, we use a cross-sectional OLS model to examine the determinants of VMT at the census tract level.

**The Ordinary Least Squares (OLS) Model**

$$\log VMT_i = \beta_0 + \beta_1 MAPC + \sum_m \delta_i^m Z_i^m + \sum_k \psi_i^k X_i^k + \varepsilon_i \quad (5)$$

Where *logVMT$_i$* is the log of average annual vehicle miles traveled by passenger vehicles in each census tract *i*; *β$_0$* is a constant term; *β$_1$* represents the impact on VMT of being a census tract in the MAPC area (compared to the other census tracts in Massachusetts). *β$_1$* will be negative if MAPC census tracts show lower daily VMT than the rest of MA. $\delta_i^m$ is the coefficient on the m-th built-environment explanatory variable. $Z_i^m$ is the value of built environment explanatory variable m for each census tract *i*. *ψ$^k_i$* is the coefficient of the *k*-th socio-economic explanatory variable. X$^k_i$ is the value of socio-demographic explanatory variable *k* for each census tract *i*. *ε$_i$* is the error term of a random variable for census tract *i*, assumed to be normally distributed with mean zero.

OLS regression, in the presence of spatial dependence, will be biased and inconsistent. To obtain estimates that are robust to any spatial dependence that is present in the variables, we then incorporate spatial lag model (SLM), spatial error model (SEM), and a mix of SLM and SEM (Anselin, 2001). We address two important issues with geographic components: spatial dependence and spatial heterogeneity (Anselin, 1988; LeSage and Pace, 2009). A spatial error model (SEM) addresses the autocorrelation among the error terms (in their neighbors) and captures the effects of omitted independent variables that are spatially correlated. A spatial lag model (SLM) addresses a variable that averages the neighboring values of a location, and it accounts for autocorrelation in the model with the weight matrix. SLM offers a possible diffusion process according to which VMT of different census tracts is affected by the explanatory variables of its own and neighboring census tracts.

**Spatial Lag Model**

$$\log VMT_i = \beta_0 + \beta_1 MAPC + \gamma W \log VMT_i + \sum_k \psi_i^k X_i^k + \sum_m \delta_i^m Z_i^m + \varepsilon_i \quad (6)$$

Where $\gamma$ is the spatial-lag correlation parameter, W is the matrix of spatial weights using longitudes and latitudes of census tract centroids.

**Spatial Error Model**

$$\log VMT_i = \beta_1 MAPC + \sum_k \psi_i^k X_i^k + \sum_m \delta_i^m Z_i^m + \varepsilon_i \quad (7)$$
$$\varepsilon_i = \lambda W \varepsilon_i + \mu_i \quad (8)$$

where $\lambda$ is an autoregressive parameter. W is the spatial weights matrix, and $\mu_i$ is the Nx1 vector of i.i.d. standard normal errors.

### 3.2.2 Sociodemographic and Built Environment Variables and Data

We use the ACS 2014 5-year estimates provided by the US Census Bureau to match with the Vehicle Census dataset. Main variables of interest include household income, age, employment status, population, and job densities. The variables derived from the Vehicle Census data are calculated as the average values for each census tract between 2009-2014. From the EPA's Smart Location Database (SLD), we obtain comprehensive built-environment data, including indicators such as the density of development, diversity of land use, and accessibility of destinations at the block group level (and aggregated to census tract for this study). These measures by SLD fall within the 2009-2014 period. Table 3 shows the details of the variables and data sources. Fig. 4 shows the spatial distribution of one of the important built environment variables, transportation accessibility index, ranging from 0 to 1. The higher the value, the better the accessibility.

**Table 3 Description of sociodemographic and built environment variables**

| Variable | Definition | Source | Year |
|---|---|---|---|
| **Sociodemographic** | | | |
| percapinc | Per capita income (In 2014 inflation adjusted dollars) | ACS | 2014 |
| avghhsize | Average household size | ACS | 2014 |
| o65y | Population over 65 years old | ACS | 2014 |
| avghhsize | Average household size | ACS | 2014 |
| unemployed | Unemployed civilian population in labor force 16 years and over | ACS | 2014 |
| veh_phh | Average registered passenger vehicle per household. Total passenger vehicles divided by number of households in 2010. | MAPC-VC | 2009-2014 |
| passvehage | Average passenger vehicle age | MAPC-VC | 2009-2014 |

| mpg_eff | Average fuel economy for passenger vehicles with valid mileage estimates, weighted by average daily mileage | MAPC-VC | 2009-2014 |
| --- | --- | --- | --- |
| w_carpool | Share of workers 16 y.o. and over who carpooled | ACS | 2014 |
| w_pubtrans | Share of workers 16 y.o. and over: Public Transportation (Includes Taxicab) | ACS | 2014 |
| w_bike | Share of workers 16 y.o. and over: Bicycle | ACS | 2014 |
| w_home | Share of workers 16 y.o. and over: Worked at Home | ACS | 2014 |
| **Built environment** | | | |
| w_avgcomm | Average one-way journey-to-work commute time (in minutes) | ACS | 2014 |
| popden | Population density (per sq. mile) | ACS | 2014 |
| roadnetden | Total road network density | EPA-SLD | 2012 |
| jobden | Gross employment density (per sq. mile) on unprotected land | EPA-SLD | 2012 |
| accessindex | Regional centrality index - relative accessibility of a block group compared to other block groups in the same MSA, expressed as travel time to working-age population via private automobiles. Values closer to 1 are more accessible | EPA-SLD | 2012 |
| dist | Distance (in miles) from Boston CBD | Computed | |
| MAPC_dum | Dummy variable, =1 if a census tract in MAPC Region, otherwise 0. | MassGIS | |

Note: ACS 2014 data are the five-year estimates.

**Fig. 4 Transportation Accessibility Index in Massachusetts**

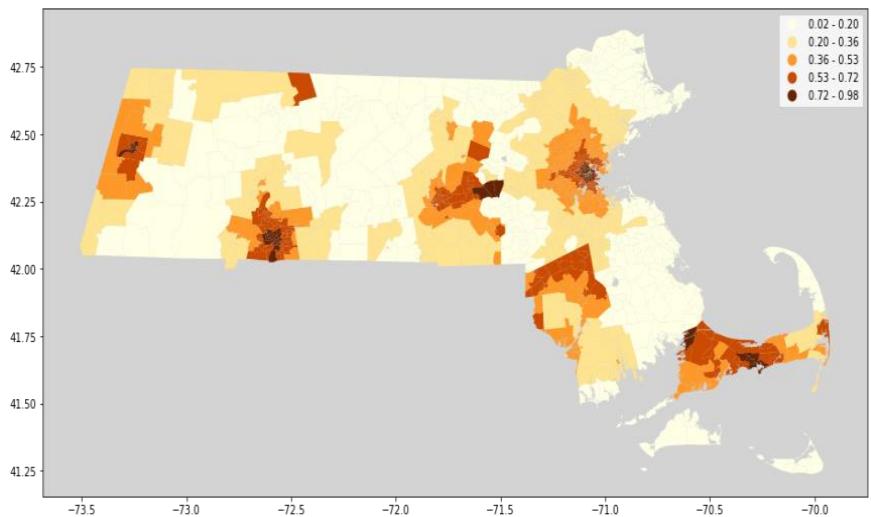

Source: EPA SLD, 2012

## 4   Results

We find that the total passenger vehicle GHG emissions estimated from the production-based perspective (i.e., using the conventional traffic-count based road inventory data) are much higher than the emissions calculated from the consumption-based standpoint (i.e., using the vehicle inspection data), as many major highways carry external traffic and exponentially higher levels of emissions. Fig. 5 shows the distribution of the total annual $CO_{2e}$ emissions using both approaches and the difference between them in the Boston Metro area.

The mean GHG emission in census tracts of the Boston metro area in 2014 is around 38,154 Mts in 2014 with the road inventory approach, while 15,624 Mts in CO2 equivalent emissions with the vehicle census approach. The median values for these two approaches are 14,946 Mts and 14,020 Mts. GHG emission estimates from the production-based (road inventory) approach show higher emissions of major road systems while neglecting the emission contributors' source. The consumption-based approach (using the vehicle odometer-reading data from the vehicle census) in assessing GHG emissions from private cars, on the other hand, allows cities to identify the strong link between household consumption and climate change and to make efforts of climate action planning through supporting the shift in consumption of travel with lower emissions. These results are particularly important when considering the GHG emissions in the context of "consumer cities" (Glaeser et al. 2001).

To understand the driving factors of VMT and to develop its mitigation strategies, we apply various econometric models to measure the relationship between VMT, socioeconomic, and built environment variables. We applied ordinary least squares (OLS), fixed effects (FE), generalized spatial lagged endogenous two-stage least squares (GS2SLS), and maximum likelihood (ML) regressors, as well as generalized methods of moments (GMM) and ML spatial error models. In Table 4, we present the goodness-of-fit of these seven models by calculating the mean squared errors (MSE) and the adjusted R-squared values. We find that the ML Spatial Lag model (SLM) has the lowest MSE and highest adjusted R-squared estimates, followed by GMM Spatial Error and Lag model (SELM) and GM2SLS SLM model. While the FE model provides similar results to spatial error and spatial lag models, spatial models, in general, are more robust.

In Table 5, we present the results from the following models: OLS, (ML) SLM, (GM) SEM, and (GMM) SELM. There are 1454 observations, each corresponding to a census tract in Massachusetts. We find that transport accessibility to destinations and road network density are associated with VMT as strongly as demographic factors. We find a significant inverse relationship between the land use measures of population, employment and road densities, and vehicle miles traveled while controlling for the effects of household size, vehicle ownership, and income.

**Fig. 5 Vehicle Emissions (Megatons of $CO_2e$) by Census Tracts in Boston Metro Area**

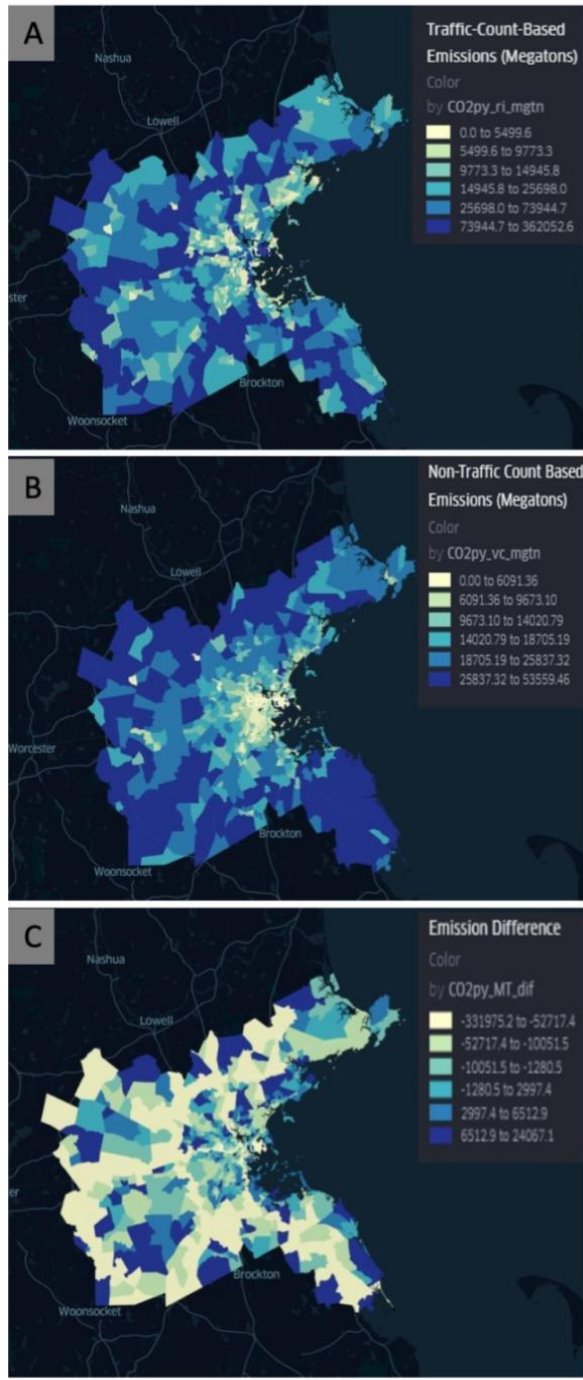

Note: (A) Traffic-count based, (B) non-traffic-count based, and (C) difference between (A) and (B).

**Table 4 Model comparison**

| Econometric Models | MSE | Adjusted R-Squared |
|---|---|---|
| OLS | 0.0048 | 0.729 |
| OLS Fixed Effects | 0.0030 | 0.838 |
| ML Spatial Lag | 0.0016 | 0.909 |
| GS2SLS Spatial Lag | 0.0045 | 0.905 |
| GMM Spatial Error | 0.0057 | 0.694 |
| ML Spatial Error | 0.0076 | 0.592 |
| GMM Spatial Error and Lag | 0.0017 | 0.904 |

The MAPC region (i.e., the Boston metro area) includes Massachusetts's inner core communities, characterized by high density, mixed-use, little space for growth, and maturing suburbs of lower density limited vacant land for new development. Places in the MAPC region are more connected, and passenger VMT is significantly lower compared to other regions in Massachusetts. Working-from-home or commuting with greener transportation modes helps reduce travel demand more in the MAPC region compared to other areas of Massachusetts.

**Table 5 Regression results for factors influencing VMT**

| Variables | OLS | (ML)SLM | (GM)SEM | (GMM)SELM |
|---|---|---|---|---|
| **Sociodemographic** | | | | |
| Log of Per Capita Income | 0.178 | 0.162** | 0.150 | 0.173** |
|  | (0.142) | (0.083) | (0.097) | (0.085) |
| Log of Per Capita Income Squared | -0.009 | -0.008* | -0.007 | -0.008** |
|  | (0.007) | (0.004) | (0.005) | (0.004) |
| % Population 65+ | -0.447*** | -0.317*** | -0.415*** | -0.352** |
|  | (0.042) | (0.025) | (0.030) | (0.025) |
| Average Household Size | 0.046*** | 0.019*** | 0.034*** | 0.021*** |
|  | (0.008) | (0.005) | (0.007) | (0.005) |
| Unemployment Rate | 0.064 | -0.003 | -0.007 | -0.003 |
|  | (0.059) | (0.035) | (0.004) | (0.036) |
| Vehicles Per Households | -0.069*** | -0.033*** | -0.038*** | -0.035*** |
|  | (0.009) | (0.006) | (0.007) | (0.006) |
| Vehicle Age | -0.027*** | -0.009*** | -0.019*** | -0.010*** |
|  | (0.004) | (0.002) | (0.003) | (0.002) |
| Fuel Efficiency (MPG) | 0.021*** | 0.009*** | 0.018*** | 0.011*** |
|  | (0.004) | (0.002) | (0.004) | (0.002) |
| % Carpooled to Work | -0.025 | 0.017 | -0.008 | 0.013 |
|  | (0.059) | (0.034) | (0.041) | (0.036) |
| % Transit to Work | -0.200** | -0.104*** | -0.144*** | -0.116*** |
|  | (0.069) | (0.040) | (0.049) | (0.042) |
| % Biked to Work | -1.256*** | -0.451** | 0.021 | -0.370** |
|  | (0.311) | (0.181) | (0.215) | (0.188) |
| % Worked from home | 0.088 (0.087) | 0.030 (0.051) | 0.134** (0.061) | 0.043 (0.053) |
| Average Commute Time | 0.011*** (0.001) | 0.003*** (0.000) | 0.006*** (0.000) | 0.004*** (0.000) |
| **Built environment** | | | | |
| Log Population Density | -0.011*** (0.001) | -0.008*** (0.002) | -0.006** (0.003) | -0.007*** (0.002) |
| Log Road Network Density | -0.044*** (0.008) | -0.018*** (0.005) | -0.029*** (0.005) | -0.019*** (0.005) |
| Log Job Density | 0.002 (0.002) | -0.004*** (0.001) | -0.001 (0.002) | -0.004*** (0.001) |
| Transportation Accessibility | -0.051*** (0.013) | -0.012* (0.007) | -0.097*** (0.015) | -0.017* (0.008) |
| MAPC Dummy | -0.046*** (0.013) | -0.011** (0.007) | -0.034*** (0.011) | -0.013** (0.008) |
| % Carpooled in MAPC | -0.300*** (0.089) | -0.135*** (0.053) | -0.135** (0.065) | -0.140*** (0.054) |
| % Biked in MAPC | -0.269 (0.335) | 0.054 (0.196) | -0.498** (0.243) | -0.047 (0.201) |
| % Transit to Work in MAPC | -0.228*** (0.069) | -0.034 (0.041) | -0.114** (0.053) | -0.043 (0.043) |
| % Worked from Home in MAPC | -0.420*** (0.122) | -0.188*** (0.071) | -0.353*** (0.089) | -0.219*** (0.074) |
| Log Distance from CBD | 0.002 (0.004) | 0.002 (0.002) | 0.019*** (0.005) | 0.004 (0.003) |
| Weighted VMT | | 0.679*** (0.013) | | 0.616*** (0.025) |
| Lambda | | | 0.547 | 0.121 |
| Constant | 8.205*** (0.740) | 2.106*** (0.442) | 8.275*** (0.500) | 2.605*** (0.497) |
| Observations | 1454 | 1454 | 1454 | 1454 |
| Adjusted (Pseudo) R-Squared | 0.734 | 0.909 | 0.694 | 0.904 |

Note: *, **, and *** denote the coefficient significant at the 0.10, 0.05 and 0.01 level respectively.

After controlling for regional proximity (measured by the distance between the studied census tract and Boston CBD) and other built environmental variables, statistically significant result for the MAPC dummy variable indicates that the plausible policies by the MAPC region to promote sustainable transportation and infrastructure have potentially played a positive role in reducing VMT. These findings support previous literature findings that justify region-specific smart-growth policies to reduce vehicle usage effectively.

Based on the mean statistics and coefficients of the socioeconomic and built environment variables presented in Table 5, we conduct a sensitivity analysis for each town or region in MA. We find that a 1% increase in the use of carpooling, public transportation, biking or working from home in the Boston metropolitan area (i.e., the MAPC region) would lead to a reduction of annual VMT by 0.13%, 0.15%, 0.41%, and 0.17% respectively, according to the SLM estimates; by 0.18%, 0.29%, 0.51%, and 0.25% respectively based on the SEM results; and by 0.14%, 0.17%, 0.43% and 0.19% respectively based on the SELM results.

Similarly, we calculate changes in annual VMT per passenger car if multiple explanatory variables change. For example, in 2014, the City of Somerville has the following values in its mean journey-to-work mode shares of carpooling (7.5%), transit (29.9%), biking (5.2%) and work-from-home (4.6%), and average log population density (9.8), log job density (8.3), log road density (3.4) and the regional accessibility index (0.69). By doubling the shares of the commute modes of carpooling, transit, biking to work, and work-from-home to 15%, 59.8%, 10.4%, and 8.6%, doubling the density variables to 19.6, 16.6, and 6.4, and increasing the regional accessibility by personal vehicles to 0.9, it would reduce annual VMT per passenger vehicle by 14.65% when using the coefficients from the SEM estimates.

## 5 Discussion

Findings on transportation emissions in this study can help understand and compare the magnitude of emitted GHGs from two different perspectives. The first one, coming from a consumption-based perspective, uses the total number of miles driven by every car owner in each census tract multiplied by the corresponding emission factors. The second one, deriving from a production-based perspective, uses an average annual daily traffic data to calculate VMT at the road segment level. We find that emissions are significantly higher for the latter approach as VMT estimates of the highway monitoring systems for all roads may include external traffic passing through the region.

We acknowledge several challenges of the study due to data limitations. First, we employed the 2014 vehicle census data from Massachusetts. Although it is the latest publicly available dataset released by the regional planning agency, it is a bit out of date as of today. Due to resource limitations, it has been challenging for planning agencies to clean, anonymize, standardize, aggregate, and publish timely large-scale data, limiting the study's time-sensitiveness. In the past few years, with technology innovation, electric vehicle consumption expansion has been a significant change for the US vehicle market. In Massachusetts, the EV numbers have increased noticeably since the launch of the MA Offers Rebates for Electric Vehicles (MOR-EV) subsidy program. In the last quarter of 2014, 795 were EVs, and 74,048 were hybrid vehicles among the over four million passenger vehicles in MA. The number of EV rebates claimed in MA has increased from under a thousand in 2014 to its highest number, around 7,000 in 2018 (MA Department of Energy Resources 2020). Therefore, the data employed in this study may not fully reflect the latest trend of passenger vehicle fleet composition in Massachusetts.

Second, the analyses presented in this paper focus on on-road passenger vehicle emissions, which only include the running exhaust $CO_2$ emissions for passenger vehicles. Emissions from other vehicle types or other processes (such as idle exhaust, start-exhaust tailpipe emissions, brake wear, tire wear, evaporative permeation, and evaporative fuel vapor venting emissions) were not included in this study. We recognize that it may underestimate GHG emissions by passenger vehicle travels, even though the running exhaust CO2 emissions contribute significantly to the vehicle emissions. For future analyses, we suggest creating a full profile of emissions that may need disaggregated vehicle trip-level or parking-level microdata (Caicedo 2010), which are unfortunately not available for this study.

Third, given the constraint that Massachusetts (MA) does not have or publish its own emission factor (EF) database, our analyses of GHG emissions by passenger vehicles are based on the EFs derived from the California EMFAC database (California Air Resources Board 2014), assuming that the fleet composition and vehicle travel speed distribution in the two states are very similar. To further improve the MA GHG emission estimation accuracy, future efforts are needed to update the emission factors that reflect the local vehicle fleet composition and travel speed distribution.

## 6  Conclusion

It is crucial to enabling city leaders and other stakeholders to take effective action to mitigate climate change by promoting sustainable development, increasing climate resilience, and reducing emissions (AECOM 2016). The application of the methodology discussed in this study to the state of Massachusetts and, more specifically, the Boston metropolitan area results in findings that have significant policy implications. Reducing VMT has been established as a critical approach to reducing local and regional GHG emissions and resolving climate change and automobile dependency (Frank et al. 2011). The transportation sector has become the fastest-growing sector in Massachusetts's carbon-related emissions over the last decade, making it crucial to address challenges within the industry. Reducing the number of driving miles will help the Commonwealth get closer to its commitment to curbing global warming, as set forth by the Massachusetts Global Warming Solutions Act of 2008 (Holloway et al. 2017).

From the analysis in this study, we suggest that the non-traffic-count based emission results from passenger vehicles in the Boston metropolitan area can be used to fund projects to target carbon-intensive consumption, to provide information through outreach campaigns that help the public shift driving activities to those with a lower level of climate impacts, to implement local government level climate action strategies including partnering with different organizations that promote sustainable consumption (USDN 2018). It can help local governments better understand their residents' driving behavior and serve as an additional point of reference both in production- and consumption-based GHG inventories. For example, the Climate Forward Action Plan for the City of Somerville aims to develop one but was advised against it. The city could use the method described here to evaluate transportation emissions at the city's census tract level.

Furthermore, the relationship between VMT and the built environment characteristics suggest the potential efficacy for reducing vehicle emissions through the configuration and siting of future development and transportation network improvements within the Boston region. One of our findings shows that density variables play a significant role in reducing VMT and GHG emissions. Increasing population, job, or road densities are generally associated with zoning changes, which is the most potent land-use management tool to promote public health, safety, and welfare (Mass Audubon 2019). Under the Massachusetts Zoning Act, municipalities have the power to establish local zoning laws that can divide communities into different zones and establish requirements for minimum lot size, street densities, open space, parking, and the process for site plan approval. For example, in December of 2019, the City Council of Somerville has approved a new zoning ordinance that eliminates mandatory parking requirements across most of the city, facilitates more transit-accessible development, and provides incentives for higher-density affordable housing (MilNeil 2019). Although a powerful planning tool, land use changes can usually take a very long time to plan and implement as they may involve new development or redevelopment, which can sometimes be impossible, especially for areas with heritage conservation regulations.

Policy incentives that motivate changes in travel behavior, vehicle types by energy sources, and energy consumption can be more effective in the short term. For example, a randomized control trial (RCT) by a large employer in the Boston region demonstrated that combining the right incentives (such as subsidized transit-pass, increased parking cost, and information for alternative travel modes) can help shift travelers' desire to reduce their carbon footprint (Rosenfield et al. 2019). Improving infrastructure and level of services for alternative modes such as bikeshare programs, and EV charging station planning (Asensio et al. 2020) can also help improve accessibility within the metropolitan area. Somerville is the first municipality in the Boston metro area to adopt changes to improve transportation options available to travelers, promote the use of efficient travel modes, reduce vehicle trips, total VMT, traffic congestion in the city, and decrease vehicle air pollutant emissions (City of Somerville 2019). Other cities and towns should learn from Somerville's experience.

In addition to using greener modes of transportation for commuting to work, telecommuting (i.e., working remotely from home) has become increasingly plausible and has been advocated by many organizations (especially in the context of growing public health concerns). Telecommuting could significantly reduce traffic and VMT, and therefore mitigate GHG emissions if applied at large scales. We find that census tracts with higher rates of telecommuters in the MAPC region have significantly lower VMT estimates. In Massachusetts, only 5.1% of the population worked remotely in 2018, with the highest rates in Newton, Worcester, and Cambridge at 10%, 6.4%, and 6.2%, respectively (FlexJobs 2018). Given the COVID-19 pandemic and the statewide lockdowns since March 2020, many private and public organizations have pledged to keep their employees working remotely for at least until the end of 2020 (Solis 2020).

A recent survey by A Better City and the City of Boston (2020) examines the commuting choices before, during, and anticipating post-pandemic based on 4200 responses from employees working in Boston. Responses to categories of telecommuting every day, a few times per month, and no-telecommuting at all, were 7%, 31%, and 50% before the pandemic; 60%, 17%, and 15% during the pandemic; and 21%, 63%, and 9% anticipating post-pandemic. An increase of 15% responded to the driving-alone mode and a slight increase of 3.5% and 1.7% for biking and walking, expecting post-pandemic. More than half of the transit users suggested that they plan to return to public transit, while the rest remain uncertain. Over 10% of all survey respondents would like to see changes in either additional dedicated bike lanes, dedicated off-road paths, or prioritized road space for bikes post-pandemic (A Better City and the City of Boston 2020).

It suggests both opportunities and challenges for planners to improve sustainable transport policies and infrastructure to reshape commuters' travel preferences post-pandemic. On the one hand, telecommuting policies by various sectors could help cities reduce VMT and GHG emissions; on the other hand, we may see a trend of rising vehicle travel. Cities have to adopt a multitude of policies and incentives to integrate land use and transportation planning, displace gasoline and diesel fuels, and motivate a shift towards electric cars and fleets to achieve their climate goals of carbon neutrality after the pandemic.

21

23